\documentclass[10pt]{wlscirep}
\usepackage{multirow}
\usepackage[utf8]{inputenc}
\usepackage[T1]{fontenc}
\usepackage{xurl}
\usepackage{hyperref}
\usepackage[font=small,labelfont=bf]{caption}
\usepackage{caption}
\usepackage{graphicx}

\title{Metagenome assembly of high-fidelity long reads with hifiasm-meta}
\author[1,2]{Xiaowen Feng}
\author[1,2]{Haoyu Cheng}
\author[3]{Daniel Portik}
\author[1,2,*]{Heng Li}
\affil[1]{Department of Data Sciences, Dana-Farber Cancer Institute, Boston, MA, USA}
\affil[2]{Department of Biomedical Informatics, Harvard Medical School, Boston, MA, USA}
\affil[3]{Pacific Biosciences, Menlo Park, CA, USA}
\affil[*]{To whom correspondence should be addressed: hli@ds.dfci.harvard.edu}

\begin{abstract}
Current metagenome assemblers developed for short sequence reads or noisy long reads
were not optimized for accurate long reads.
Here we describe hifiasm-meta, a new metagenome assembler that exploits the high accuracy of recent data.
Evaluated on seven empirical datasets, hifiasm-meta reconstructed tens to hundreds of complete circular bacterial genomes per dataset,
consistently outperforming other metagenome assemblers.
\end{abstract}

\begin{document}     
\maketitle

\noindent \emph{De novo} assembly of metagenome samples is a common approach to the study of microbial communities~\cite{Lapidus:2021vb}.
Contigs in a metagenome assembly of short reads are usually tens of kilobases (kb) in length~\cite{almeida2019new}, $\sim$1\% of a bacterial genome.
After years of metagenome sequencing, there were only 62 complete genomes assembled from metagenome samples as of September 2019~\cite{Chen:2020aa}.
Although we can cluster short contigs into metagenome-assembled genomes (MAGs) with binning algorithms~\cite{kang2019metabat},
binning can be an important source of errors which complicate or mislead downstream analysis~\cite{Chen:2020aa}.
The limitation of short-read MAGs motivated the development of metaFlye~\cite{Kolmogorov2020-gu},
the only published assembler specialized for long-read metagenome assembly.
Initially developed for noisy long reads of error rate $\sim$10\%, Flye~\cite{Kolmogorov:2019aa}, which metaFlye is based on,
does not take advantage of PacBio's high-fidelity reads (HiFi) and is suboptimal for single-species HiFi assembly~\cite{Nurk2020-zh}.
To leverage the full power of long accurate HiFi reads, we developed hifiasm-meta, extending our earlier work~\cite{Cheng2021-np} to metagenome samples.

In comparison to the assembly of a single species, metagenome assembly poses several unique challenges~\cite{Lapidus:2021vb,cao2021reconstruction}, such as
a larger variance in read length distribution in PacBio HiFi data,
and high ploidy combined with low coverage in certain haplotypes.
We made several major changes in hifiasm-meta to address these challenges.
First, hifiasm-meta has an optional read selection step that reduces the coverage of highly abundant strains without losing reads on low abundant strains.
Second, during the construction of the assembly graph, hifiasm-meta tries to protect reads in genomes of low coverage, which may be treated as chimeric reads and dropped by the original hifiasm.
Third, hifiasm-meta only drops a contained read if other reads exactly overlapping with the read are inferred to come from the same haplotype.
This reduces contig breakpoints caused by contained reads~\cite{DBLP:conf/isit/HuiSRC16}.
Fourth, after the initial graph construction, hifiasm-meta uses the coverage information to prune unitig overlaps,
assuming unitigs from the same strain tend to have similar coverage.
It also tries to join unitigs from different haplotypes to patch the remaining assembly gaps.
These strategies make hifiasm-meta more robust to features in metagenome datasets.

We first evaluated hifiasm-meta,
metaFlye~\cite{Kolmogorov2020-gu} and HiCanu~\cite{Nurk2020-zh}
on two mock communities ATCC and zymo (Table~1).
ATCC consists of 20 distinct species, 15 of which are abundant at 0.18--18\% and 5 are rare at 0.02\% abundance.
We were able to reconstruct 14 of the abundant species each as a complete circular contig,
a slight improvement over metaFlye and Hicanu (Table~S1).
All tools assembled \emph{P. gingivalis}, at 18\% abundance, into two contigs.
No assemblers could fully reconstruct the five species of low abundance.
We manually checked the read alignment of these species and found their assembly gaps are all caused by insufficient coverage.
We would not be able to assemble these species in full with the current data.
The zymo dataset features 21 strains of 17 species, including five strains of \emph{E. coli} at 8\% abundance each.
A challenge of this dataset lies in the phasing of the \emph{E. coli} strains.
Hifiasm-meta assembled strain B766 into a complete circular contig, 
strain B3008 into 2 contigs and the rest as fragmented contigs.
HiCanu assembled both B766 and B3008 into complete circular contigs;
metaFlye failed to assemble all 5 strains as circular contigs. 
Hifiasm-meta produces a more contiguous assembly for \emph{M. smithii} at 0.04\% abundance (Table~S1).
Generally, all three assemblers have comparable accuracy on the two mock community datasets.

\begin{table}[tb]
\caption{Evaluated metagenome datasets}
{
\begin{tabular*}{\textwidth}{@{\extracolsep{\fill}} llrrrl}
\hline
\multirow{2}{*}{Sample} & \multirow{2}{*}{Accession} & {\# bases} & {N50 read} & {Median} & \multirow{2}{*}{Sample description}\\
& & (Gb) & length (kb) & read QV &\\
\hline
ATCC        & SRR11606871 & 59.2         & 12.0                & 36              & Mock community ATCC MSA-1003~\cite{Hon2020-ve}\\
zymo        & SRR13128014 & 18.0         & 10.6                & 40              & Mock community ZymoBIOMICS D6331\\
sheepA      & SRR10963010 & 51.9         & 14.3                & 25              & Sheep gut microbiome~\cite{Kolmogorov2020-gu}\\
sheepB      & SRR14289618 & 206.4        & 11.8                & N/A              & Sheep gut microbiome~\cite{Bickhart2021.05.04.442591}\\
humanO1     & SRR15275213 & 18.5         & 11.4                & 40              & Human gut from a pool of four omnivore samples\\
humanO2     & SRR15275212 & 15.5         & 10.3                & 41              & Human gut from a pool of four omnivore samples\\
humanV1     & SRR15275211 & 18.8         & 11.0                & 39              & Human gut from a pool of four vegan samples\\
humanV2     & SRR15275210 & 15.2         & 9.6                 & 40              & Human gut from a pool of four vegan samples\\
chicken     & SRR15214153 & 33.6         & 17.6                & 30              & Chicken gut microbiome \\
\hline
\end{tabular*}
}
\begin{flushleft}
\footnotesize The N50 read length is the length of the shortest read at 50\% of the total number of read bases.
The quality value (QV) of a read is $-10\log_{10}e$, where $e$ is the expected sequencing error rate of the read, assuming accurate base quality.
No base quality is available for the sheepB dataset.
\end{flushleft}
\end{table}

\begin{figure}[ph]
\centering
\includegraphics[width=0.95\textwidth]{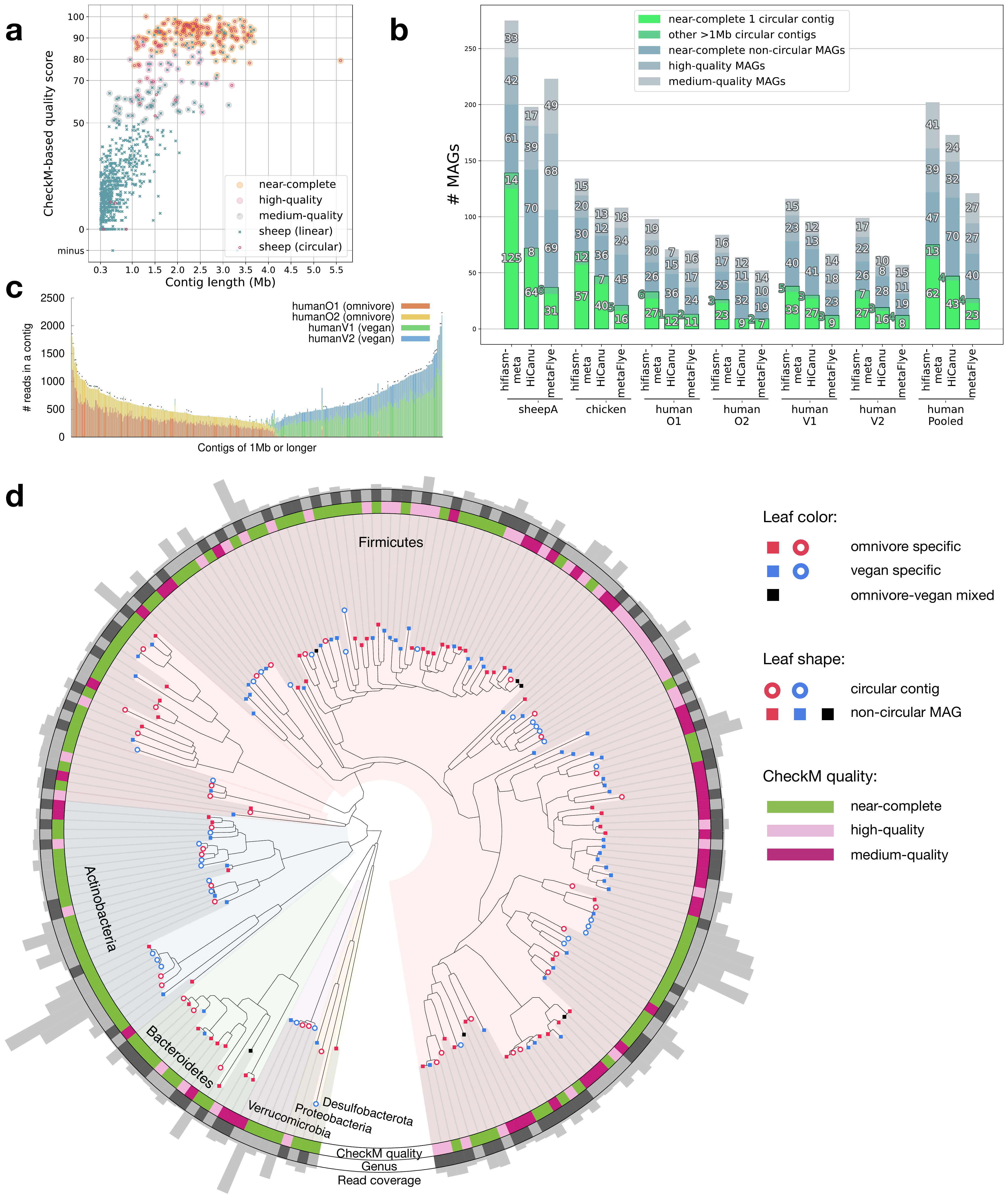}
\caption{Metagenome assemblies of empirical datasets.
(a) Quality score of contigs longer than 300kb from the hifiasm-meta assembly of sheepA.
The quality score of a MAG is defined as ``${\rm completeness}-5\times{\rm contamination}$'' based on CheckM reports.
(b) CheckM evaluation.
A MAG is ``near-complete'' if its CheckM completeness is $\ge$90\% and its contamination level $\le$5\%,
is ``high-quality'' if completeness $\ge$70\% and contamination $\le$10\%,
or is ``medium-quality'' if its quality score is $\ge$50\%.
``HumanPooled'' represents the co-assembly of all four human gut samples.
(c) Composition of long contigs in the hifiasm-meta co-assembly of four human gut samples.
Each bar represents a contig of $\ge$1Mb in length.
It gives the number of reads used in the contig.
Each color corresponds to a human gut sample.
A cross at the top of a bar indicates the contig being circular.
(d) Phylogeny of human gut MAGs from the co-assembly.
A colored clade corresponds to a phylum inferred by GTDB-Tk~\cite{Chaumeil:2019uz,asnicar2015compact}.
A MAG is omnivore/vegan-specific if 90\% of reads in the MAG come from omnivore/vegan samples, respectively.
Two adjacent leaves coming in the same genus have the same shade on the ``Genus'' outer ring.
}
\end{figure}

We then evaluated the three HiFi metagenome assemblers on real datasets (Table~1).
Due to the lack of their true compositions, we used CheckM~\cite{Parks2015ua} to measure
the completeness and the contamination level of each assembly.
From the sheepA gut sample, hifiasm-meta reconstructed 328 contigs longer than 1Mb (Fig.~1a; Extended Data Fig.~1) totaling 656Mb in length.
173 of them were near-complete according to CheckM (Fig.~1b).
Most long contigs that failed to reach this category are due to incompleteness, not due to excessive contamination.
Among the 173 near-complete hifiasm-meta contigs, 125 are circular (Fig.~1b),
representing a significant improvement over 
HiCanu (64 circular near-complete contigs) and metaFlye (31).
We aligned hifiasm-meta, HiCanu and metaFlye assemblies to each other and investigated the similarity between them.
We found 86\% and 72\% of circular near-complete HiCanu and metaFlye contigs, respectively, are also circular in the hifiasm-meta assembly and are of similar lengths (Table~S2).
The remaining near-complete circular HiCanu and metaFlye contigs are assembled into either one linear contig or two linear contigs by hifiasm-meta.
Hifiasm-meta can reconstruct most high-quality contigs found by other assemblers.
Furthermore, the mash~\cite{ondov2016mash} sequence divergence between hifiasm circular contigs is above 1\% for all contig pairs.
This suggests hifiasm-meta is recovering more species or strains in full and the higher number of circular contigs is not due to redundancies in our assembly.

To reconstruct MAGs from non-circular contigs, we applied the MetaBAT2 binning algorithm~\cite{kang2019metabat} to each assembly.
Not optimized for long-read assemblies, MetaBAT2 may mistakenly group different strains of the same species into one MAG
and even group two circular contigs occasionally.
Such MAGs would be considered to be contaminated by CheckM.
To improve binning, we separated circular contigs into individual bins.
In the end, we identified more than $>$120 non-circular MAGs of medium or higher quality from each sheepA assembly (Fig.~1c).
Hifiasm-meta still finds more quality MAGs in total.

We applied hifiasm-meta to the larger sheepB dataset~\cite{Bickhart2021.05.04.442591} (Table~1)
and obtained 438 near complete MAGs and 245 circular contigs.
Bickhart et al~\cite{Bickhart2021.05.04.442591} assembled the combined sheepA and sheepB datasets with metaFlye and clustered contigs into MAGs using additional Hi-C data.
They reported 44 circular contigs and 428 near complete MAGs evaluated by DAS Tool~\cite{sieber2018recovery}.
For a direct comparison, we ran CheckM on their assembly and identified 241 near complete MAGs instead.
Hifiasm-meta produced a more contiguous assembly with HiFi data only.

For the chicken and the four human gut metagenomes (Table~1), hifiasm-meta consistently produced more circular contigs and more total MAGs than HiCanu and metaFlye as well (Fig.~1c).
All assemblers produced fewer MAGs in comparison to the sheepA gut sample.
To see how much this is caused by the higher data volume of sheepA,
we randomly sampled $\sim$18Gb of sequences from this dataset, comparable to the size of humanO1.
On the downsampled dataset, we could assemble 84 circular contigs, doubling the number of circular contigs in humanO1.
This suggests that data volume does affect the assembly quality but
the more contiguous sheepA assembly is probably more related to the property of the sample.

Among the four human gut datasets, two were collected from omnivore donors and the other two from vegan donors.
Each dataset represents a pool of four individuals (Table~1).
We further pooled the four datasets together and co-assembled them.
With read names reported in the final hifiasm-meta assembly,
we can identify the composition of each contig based on the sources of reads.
We found the great majority of contigs of $\ge$1Mb in size
and almost all $\ge$1Mb circular contigs are either omnivore-specific or vegan-specific (Fig.~1c),
whereas the two omnivore samples are well mixed in long omnivore-specific contigs, so it is with the two vegan samples.
We see more reads coming from the humanO1 and humanV1 datasets probably because humanO1 and humanV1 have more and longer reads than the other two human gut samples.

Omnivore and vegan samples are also well separated among co-assembled MAGs,
though omnivore- and vegan-specific MAGs are mixed in the phylogenetic tree (Fig.~1d):
in this tree, 19 genera consist of three or more MAGs, 16 of which contain both omnivore- and vegan-specific MAGs.
This suggests hifiasm-meta assembly is better at untangling subtle composition differences.
Also notably, a clade of seven circular contigs (in the northeastern direction in Fig.~1d) only has 75--79\% CheckM completeness.
Two of them were assembled by HiCanu into circular contigs of near identical lengths, so they are less likely to be truncated misassembly.
We speculate this clade may be underrepresented in CheckM.

On performance, hifiasm-meta took $\sim$18 hours over 48 CPU threads to assemble the sheepA and the chicken datasets
and took $\sim$3 hours for the human gut samples (Table~S3).
On these datasets, it is as fast as metaFlye and is consistently faster than HiCanu by a few folds.
Hifiasm-meta tends to use more memory than metaFlye and HiCanu, consuming $\sim$200Gb RAM for the sheepA and chicken gut samples.
Hifiasm-meta assembled the largest sheepB dataset in 8.9 days and used 724Gb RAM at the peak.

In the era of short-read sequencing, metagenome assembly was rarely considered as a method to reconstruct full genomes~\cite{Chen:2020aa}.
This view has been changed by recent progress in long-read assembly~\cite{Moss:2020vt,Vicedomini:2021tl,Kolmogorov2020-gu,Bickhart2021.05.04.442591}.
Optimized for long accurate HiFi reads, hifiasm-meta moves metagenome assembly even further.
It possibly assembles more circular MAGs from one deeply sequenced sample, without manual intervention, than all circular MAGs published in the past.
Such high-quality metagenome assemblies may fundamentally change the practice in metagenome analysis and shed light on the biological and biomedical implications of microbial communities.

\section*{Methods}

\noindent{\bf Overview of the hifiasm-meta algorithm.}
The hifiasm-meta workflow consists of optional read selection, sequencing error correction,
read overlapping, string graph construction and graph cleaning.
The error correction and read overlapping steps are largely identical to the original hifiasm.
We added optional read selection and revamped the rest of steps.

~\\
\noindent{\bf Optional downsampling of input reads.}
If read selection is enabled, hifiasm-meta will first make a crude guess of 
whether there are too many alignments to be performed for the whole read set. 
This is done by examining anchors and is alignment free.
We proceed to do the selection if 2/3 reads have more than 300 target reads.
We start with an empty hash table which will record kmer counts, 
and go through reads in batches of 2000.
In a batch, for each read encountered, 
we collect its canonical kmers and query the hash table for their occurrences.
Three percentiles 3\%, 5\% and 10\% are checked against 
the corresponding thresholds 10, 50 and 50 respectively.
If any percentile is lower than the given threshold, the read is kept.
The rationale is that we would like to keep a read when it has some rare kmers,
i.e. when discarding it will lead to loss of information.
Note that the ``rare kmers'' here are not necessarily rare globally, 
and the read selection result might change if the inputs are shuffled.
We assume that the input is not particularly sorted.
After all reads in the batch have been processed, 
we update the kmer counting hash table with them (kmers of discarded reads are also counted).
The termination criterion of the read selection is 
either the total number of reads being kept has exceed the desired count,
or all reads have been processed.

~\\
\noindent{\bf Modified chimera detection.}
Before graph construction, the original hifiasm regards a read to be chimeric and discards it if a middle part of the read is not covered by other reads.
A read from a genome of low abundance may have such an uncovered region due to statistical fluctuation.
Hifiasm-meta disables the heuristic if both ends of the read overlap with five or fewer other reads.
This extra threshold improves the contiguity of genomes of low abundance.

~\\
\noindent{\bf Treatment of contained reads.}
The standard procedure to construct a string graph discards a read contained in a longer read.
This may lead to an assembly gap if the contained read and the longer read actually reside on different haplotypes~\cite{DBLP:conf/isit/HuiSRC16}.
The original hifiasm patches such gaps by rescuing contained reads after graph construction.
Hifiasm-meta tries to resolve the issue before graph construction instead.
It retains a contained read if other reads exactly overlapping with the read are inferred to come from different haplotypes.
In other words, hifiasm-meta only drops a contained read if there are no other similar haplotypes around it.
This strategy often retains extra contained reads that are actually redundant.
These extra reads usually lead to bubble-like subgraphs and are later removed by the bubble popping algorithm in the original hifiasm.

~\\
\noindent{\bf Changes to graph cleaning.}
At the graph construction stage, the original hifiasm-meta rejects overlaps between unitigs inferred to come from different haplotypes.
Hifiasm-meta may do this to patch remaining assembly gaps.
Hifiasm-meta also uses the unitig coverage to prune overlaps.
Suppose unitig $A$ overlaps unitig $B$ and $C$ in the same orientation.
Such a bifurcation is an ambiguity in the assembly graph.
Let $r_{AB}=\min\{{\rm cov}(A),{\rm cov}(B)\}$, where ${\rm cov}(A)$ is the coverage of $A$.
Hifiasm-meta drops the overlap between $A$ and $C$ if $r_{AB}>0.7$ and $r_{AC}<0.7$.
This strategy is only applied to unitigs longer than 100kb as it is difficult to accurately estimate coverage for short unitigs.
In addition, attempt to resolve short unitigs would not greatly improve the assembly quality anyway in our testing.

~\\
\noindent{\bf Assembly of metagenome datasets.}
We evaluated hifiasm-meta r52, HiCanu v2.1.1 and metaFlye v2.8.1 all with 48 CPU threads.
We used ``{\tt hifiasm-meta reads.fa}'' for the assembly of empirical gut samples
and used ``{\tt hifiasm-meta -{}-force-rs -A reads.fa}'' to enable read selection for the two mock community datasets.
We ran HiCanu with ``{\tt canu maxInputCoverage=1000 genomeSize=100m batMemory=200 -pacbio-hifi reads.fa}''.
We tried to increase the ``{\tt genomeSize}'' parameter to 1000m for sheepA and got identical results.
We ran metaFlye with ``{\tt flye -{}-pacbio-hifi reads.fa -{}-plasmids -{}-meta}''.
Hifiasm-meta and metaFlye report assembly time and peak memory usage.
We used GNU time to measure the performance of HiCanu.

~\\
\noindent{\bf Metagenome binning.}
We used MetaBAT2 for initial binning and then post-process MetaBAT2 results to get final MAGs.
We aligned raw reads to an assembly with
``{\tt minimap2 -ak19 -w10 -I10G -g5k -r2k -{}-lj-min-ratio 0.5 -A2 -B5 -O5,56 -E4,1 -z400,50 contigs.fa reads.fa}'',
calculated the depth with ``{\tt jgi\_summa\_rsize\_bam\_contig\_depths -{}-outputDepth depth.txt input.bam}'',
and ran MetaBAT2 with
``{\tt metabat2 -{}-seed 1 -i contigs.fa -a depth.txt}''.
We tried different random seeds and got similar results.
We only applied MetaBAT2 to the primary hifiasm-meta and HiCanu assemblies as including alternate assembly led to worse binning.
After MetaBAT2 binning, we separate circular contigs of 1Mb or longer into a separate MAG if it is binned together with other contigs.

~\\
\noindent{\bf Evaluating metagenome assemblies.}
We ran CheckM v1.1.3 to measure the completeness and the contamination level of MAGs.
The command line is
``{\tt checkm lineage\_wf -x fa input/ wd/; checkm qa -o 2 wd/lineage.ms .}''.
We also tried DAS Tool for evaluation on the sheepA dataset.
DAS Tool is more optimistic, identifying 22\% more near-complete MAGs in comparison to CheckM.
As CheckM is more often used for evaluation, we only applied CheckM to all assemblies.

We used GTDB-Tk v1.3.0 for phylogenetic placement with command line
``{\tt GTDBTK\_DATA\_PATH=GTDB-Tk/release95 gtdbtk classify\_wf}''.
We annotated the tree and used GraPhlAn for visualization. 

\section*{Acknowledgements}
We thank Wei Fan for sharing the chicken gut dataset.
This study was supported by US National Institutes of Health (grant R01HG010040 and
U01HG010971 to H.L.).

\section*{Author contributions}
X.F. and H.L. conceived the project, designed the algorithm and wrote the manuscript.
X.F. implemented the algorithm and evaluated the metagenome assemblies.
All co-authors helped the data analysis and revised the manuscript.

\section*{Competing interests} 
D.P. is an employee of Pacific Biosciences.
H.L. is a consultant of Integrated DNA Technologies and on the Scientific Advisory Boards of Sentieon and Innozeen.

\section*{Data availability}
HiFi data were obtained from NCBI Sequence Read Archive (SRA) with accession numbers shown in Table~1.
All generated assemblies are available at \url{ftp://ftp.dfci.harvard.edu/pub/hli/hifiasm-meta/v1}.
ZymoBIOMICS mock reference genomes were downloaded from \url{https://s3.amazonaws.com/zymo-files/BioPool/D6331.refseq.zip}.
The list of reference genomes in the ATCC mock community is available at \url{https://www.atcc.org/products/msa-1003}.
CheckM database: \url{https://data.ace.uq.edu.au/public/CheckM\_databases/checkm\_data\_2015\_01\_16.tar.gz}.
GTDB-Tk database: \url{https://data.ace.uq.edu.au/public/gtdb/data/releases/release95/95.0/auxillary\_files/}.

\section*{Code availability}
Hifiasm-meta is available at \url{https://github.com/xfengnefx/hifiasm-meta}.

\bibliography{main}

\section*{Extended Data Figures}
    \begin{center}
        \captionsetup{type=figure}
        \includegraphics[width=0.8\textwidth]{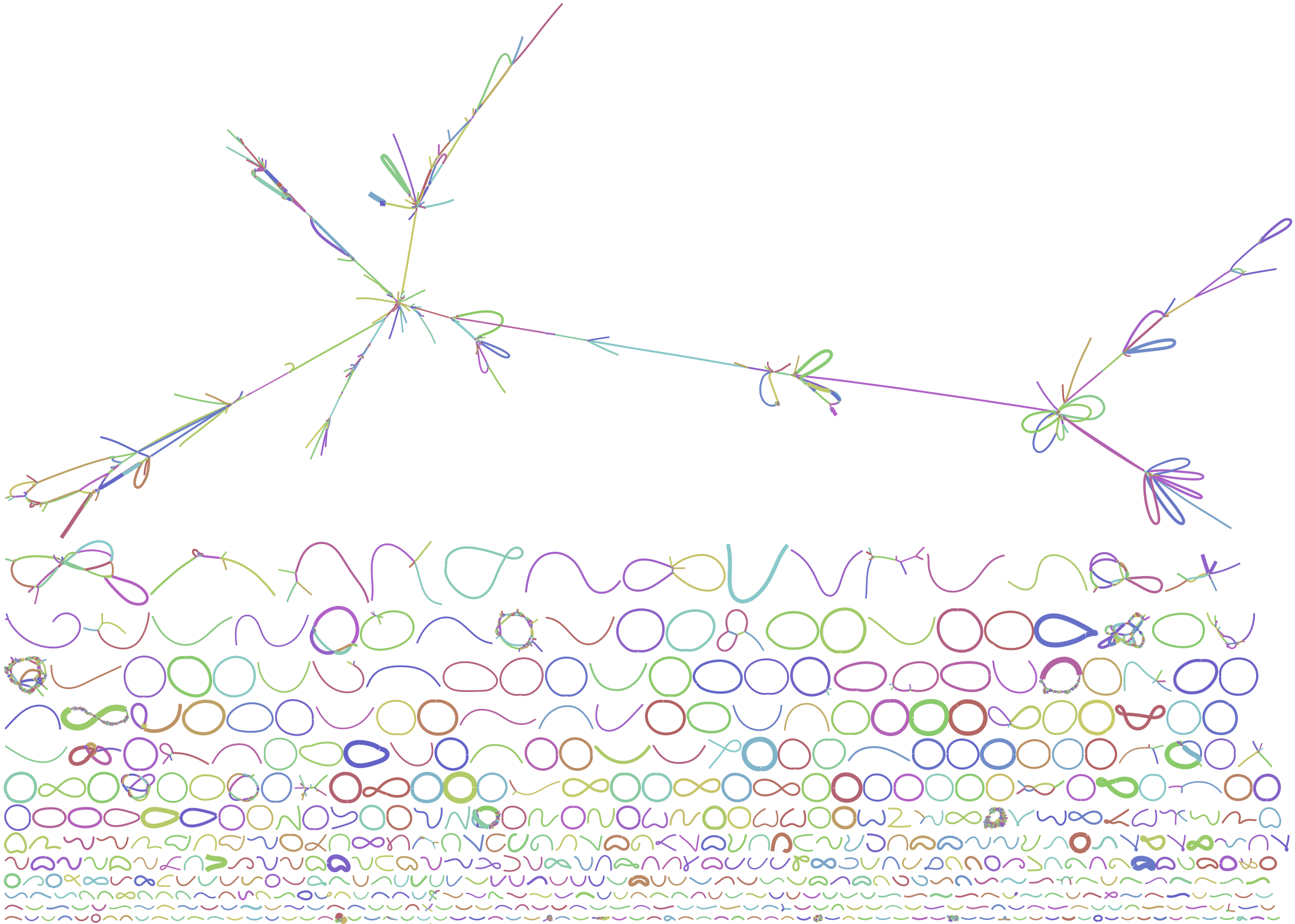}
        \captionof*{figure}{{\bf Extended Data Figure 1.} The hifiasm-meta assembly graph of the sheepA dataset. Short disconnected contigs are not shown.}
    \end{center}

\end{document}